# Providing Group Anonymity Using Wavelet Transform

Oleg Chertov[1], Dan Tavrov[1]

[1] Faculty of Applied Mathematics, National Technical University of Ukraine
"Kyiv Polytechnic Institute", 37 Peremohy Prospekt, 03056 Kyiv, Ukraine
{chertov, kmudisco}@i.ua

**Abstract.** Providing public access to unprotected digital data can pose a threat of unwanted disclosing the restricted information.

The problem of protecting such information can be divided into two main subclasses, namely, individual and group data anonymity. By group anonymity we define protecting important data patterns, distributions, and collective features which cannot be determined through analyzing individual records only.

An effective and comparatively simple way of solving group anonymity problem is doubtlessly applying wavelet transform. It's easy-to-implement, powerful enough, and might produce acceptable results if used properly.

In the paper, we present a novel method of using wavelet transform for providing group anonymity; it is gained through redistributing wavelet approximation values, along with simultaneous fixing data mean value and leaving wavelet details unchanged (or proportionally altering them). Moreover, we provide a comprehensive example to illustrate the method.

## 1 Introduction

The problem of preserving privacy has become pressing in the recent years, and this fact doesn't seem to change in the nearest future. That is mainly due to the rapid growth of digital data [1], and the enhancement of public access to collected information. The latter means that with an additional permission one can easily get access to the great variety of primary data such as information on patients' hospital treatment (so-called Clinical Data Repositories [2]), electronic commerce results in big automated collections of consumer data, microfiles with large census (or other sociological surveys) data samples etc. The most fundamental project is without a doubt IPUMS-International [3]. Within it, more than 279 million person records collected from 130 censuses held in 44 countries (at the moment this paper is being written) are accessible for the researchers. Of course, this information is totally de-identified. To preserve its privacy, special data anonymity methods need to be used. Moreover, this is often a subject to legal regulation. E.g., in the USA, to comply with the Health Insurance Portability and Accountability Act of 1996 (HIPAA) [4] and the Patient Safety and Quality Improvement Act of 2005 (PSQIA) [5], organizations and individuals don't have to reveal their medical data without preceeding privacy protection in any case. Besides, some consumer information transfer shouldn't lead to the individual persons' public profiling, and is a subject of a strict regulation. For instance, see Directive on privacy and electronic communications [6] to learn more about such regulations in the EU.

So, the overall accessible information amount growth, emerging of various (and often cross-referencing) sources to get it, developing of the data mining methods for finding out implicit data patterns, and necessity of following appropriate regulation rules issue more and more challenges to respond to before publishing the data.

Works on privacy protection in publicly accessed data can be considered as a part of privacy-preserving data mining field.

Usually, the first thing to define is microdata which mean information on respondents (e.g., persons, households, or companies). Respectively, a microfile is a set of microdata reduced to one file that consists of attributive records describing each respondent. Statistical disclosure control (SDC) methods aim at receiving new, protected microdata basing on the original ones. But, such a procedure should meet following conditions [7, p. 399]:

- Disclosure risk is low or at least adequate to protected information importance.
- Both original and protected data, when analyzed, yield close, or even equal results.
- The cost of transforming the data is acceptable.

These requirements are equivalent to an assertion that the informational and financial losses during microdata transformation have to be acceptable, and the level of disclosure risk has to be adequate. In other words, the microdata distortion should be small enough to preserve data utility but sufficient enough to prevent exposing the private information on *individuals* (or *groups of individuals*) within the released data.

We can mark out following SDC methods that are heavily used in practice:
- *randomization* – a noise is added to the data to mask records' attribute values [8];
- *microaggregation* – a set of original records is partitioned into several groups such way that the records in a group are similar. In addition, there are at least $k$ records in each group. The average value over each group is computed for every attribute. Then this value is used to replace each of the original ones [9];
- *data swapping* – transforming the microfile by exchanging values of confidential attributes among individual records [10];
- *non-perturbative methods* – protecting data without altering them. These methods are based on suppression and generalization (recoding). Suppression means removing some data from the original set, whereas recoding is data enlargement [11].

Apart from these ones, matrix decomposition [12] and factorization [13] techniques have also been used for distorting numeric sets of data (in the applications of privacy-preserving data mining). And, of course, using wavelet transform (WT) seems to be a perspective approach as well. E.g., we can use discrete WT to decompose primary data into an approximation and details with corresponding coefficients. After that, we can suppress the high-frequency detail coefficients to gain data distortion.

But, all the methods mentioned above are designed to provide data anonymity of individuals. At the same time, a problem of providing data anonymity of a respondent group remains open [7]. By *anonymity of a group* (or simply *group anonymity*) we define protecting important data distributions, patterns and features that cannot be revealed by analyzing individual records only. Providing group anonymity means performing specific primary data rearrangements which guarantee preserving privacy of a particular attribute values' distribution (for a determined respondent group). E.g., we could possibly want to protect military personnel regional distribution, or to hide the information on how the drugs are spread among different ethnic groups.

These tasks only seem to be easy-to-solve. Of course, we might swap values standing for the "Region of work" attribute between military base officers and those civilians with other attribute values similar. As a result, we would possibly conceal the base location. But, there is huge downside: such data swapping can badly influence the overall data utility.

In general, providing group anonymity implies deliberate modifying the respondents' distribution over specific attribute values. But, the data utility must be necessarily preserved. By such a utility we understand ratios between some respondent groups, or other appropriate relative values.

Let's consider one typical example. Real regional distribution of military and special service officers can be totally confidential. But, information on their distribution by age or, say, number of family members can be an interesting subject to sociological researches.

In this paper, we propose using WT to find some balance between modifying primary data and preventing loss of their utility. But, we suggest using it in a way opposite to the one applied for providing individual anonymity (see [14]). To protect data, we redistribute WT approximation values. To prevent utility loss, we fix the data mean value and leave WT details unchanged (or alter them only proportionally). In this case, ratios between various attribute values ranges will persist. Let's take [15] as an illustration. In Russia, 44 public opinion polls (1994-2001) showed that WT details actually reveal hidden time series features which are significant for near- and medium-term social processes forecasting.

The rest of this paper is arranged as follows. We provide a brief review of the related work in Section 2. The basics of our wavelet-based method for providing group anonymity are presented in Section 3. Experimental results of applying the method to a model example are discussed in Section 4. Finally, a brief conclusion is given in Section 5.

## 2 Related Work

There exist two main approaches to completing the task of protecting the confidential information. The classical one lies in encrypting the data or protecting them using different means like restricting public access. The main target of this approach is to disable (complicate) obtaining the data. In the paper, we do not consider it. We examine another one that provides SDC methods instead.

In the past decade, there has been published various literature on data anonymity (consult the anonymity bibliography from the Free Haven Project [16]).

We can divide all SDC methods into two large classes, namely, randomization methods and group-based anonymization methods.

The randomization methods [8, 17] are simple techniques which can be easily implemeted while collecting the data. It is possible because the noise being added to a record is independent of the other records' values. In addition, these methods serve well at preserving data utility, especially patterns and association rules. But, their advantages give rise to their main downsides. Provided there are other sources of information with publicly available records intersecting with the data to be masked, privacy can be violated with a great possibility [18]. In particular, this is the case with the outlier records which can easily be distinguished among the other ones in the same area [19].

A typical example of the group-based anonymization is so-called $k$-anonymity [20]. Its main idea is to ensure that every attribute values combination corresponds to at least $k$ respondents in the dataset. To achieve that, different methods can be used, mentioned in [9, 10, 11] being the most popular.

For the last 4-5 years, WT has also been used for providing data anonymity, though it has widely been used mainly in signal processing [21, 22] before. Paper [23] presents a good overview of applying wavelets to data mining in general.

Paper [24] was the first work to introduce WT into preserving data anonymity. It proposed a new data-modification algorithm for revealing data patterns without revealing the data themselves. But, the wavelet-perturbed dataset has different dimensions in the transformed space, if to compare with the original one. Later, a method [14] free of this disadvantage was introduced. In [25], the same authors improved it with simultaneous privacy- and statistics-preserving. They showed that normalizing the data guarantees the persistance of their mean value and standard deviation. Another technique proposed in [14, 25] reduces the high-frequency "noise" hidden in the original data entries by thresholding WT detail coefficients. Thus, the respondent anonymity can be achieved.

But, all these methods guarantee individual anonymity only. To solve the problem of providing group anonymity stated in [7], we introduce a novel wavelet-based method. We tend to achieve anonymity by redistributing approximation values, but we also try to save data utility by fixing the details. Figuratively speaking, we change the relief of a restricted area but try to preserve local data distribution.

## 3  Theoretic Background

### 3.1  General Definitions

Let the microfile data be organized in a table similar to Table 1. Here, $m$ stands for a number of respondents, $q$ stands for a number of attributes, $w_j$ stands for the $j^{th}$ attribute, $r_i$ stands for the $i^{th}$ record, $z_{ij}$ stands for a microfile data element.

**Table 1.** Microfile data.

|       | $w_1$    | $w_2$    | $\ldots$ | $w_q$    |
|-------|----------|----------|----------|----------|
| $r_1$ | $z_{11}$ | $z_{12}$ | $\ldots$ | $z_{1q}$ |
| $r_2$ | $z_{21}$ | $z_{22}$ | $\ldots$ | $z_{2q}$ |
| $\ldots$ | $\ldots$ | $\ldots$ | $\ldots$ | $\ldots$ |
| $r_m$ | $z_{m1}$ | $z_{m2}$ | $\ldots$ | $z_{mq}$ |

Solving group anonymity problems implies redistributing elements $z_{ij}$ according to some purpose. Let us formally set a corresponding task.

We will denote by $S_v$ a subset of a Cartesian product $w_{v_1} \times w_{v_2} \times ... \times w_{v_l}$ of Table 1 columns. Here, $v_i$, $i = \overline{1, l}$ are integers. We will call an element $s_k^{(v)} \in S_v$, $k = \overline{1, l_v}$, $l_v \le l$ *a vital value combination* because such combinations are vital for solving our task. Respectively, each element of $s_k^{(v)}$ will be called *a vital value,* and $w_{v_j}$ will be called *a vital attribute*.

Group anonymity is gained through redistributing records with specific vital value combinations. E.g., to redistribute "Middle-aged women" we need to take "Age" and "Sex" as vital attributes.

We will also denote by $S_p$ a subset of microfile data elements $z_{ip}$ corresponding to the $p^{th}$ attribute, $p \ne v_i \;\; \forall i = \overline{1, l}$. Elements $s_k^{(p)} \in S_p$, $k = \overline{1, l_p}$, $l_p \le l$ will be called *parameter values*, whereas $p^{th}$ attribute will be called *a parameter attribute*. We call it this way becaues we will use it to divide microfile records into categories.

For instance, having taken "Region" as a parameter attribute, we obtain groups of residents living in particular area.

Thus, providing group anonymity actually means redistributing records with particular vital value combinations over various parameter values.

Having defined attributes, we need to calculate the number of microfile records with a specific pair of a vital value combination and a parameter value. In many cases, absolute numbers do not provide important information on data distribution features and are not representative. Thus, modifying them can guarantee data privacy but surely leads to siginificant loss of data utility.

On the other hand, redistributing ratios sounds like a much better idea. That's why we need to divide the absolute numbers by the overall number of records in the same group. E.g., to protect "Middle-aged women", we can divide their quantity by the overall number of "Women", or "Middle-aged people", or even "People" in general (coming from a particular task to be completed). Obtained ratios can be gathered in an array $c = (c_1, c_2, ..., c_n)$ which we will call *a concentration signal*.

According to Section 1, we need to construct a new concentration signal $\tilde{c} = (\tilde{c}_1, \tilde{c}_2, ..., \tilde{c}_n)$ by redistributing the wavelet approximation of signal $c$. At the

same time, we need to preserve data utility by fixing the signal mean value
($\sum_{i=1}^{n} c_i = \sum_{i=1}^{n} \tilde{c}_i$) and wavelet details (or altering these details only proportionally).

In the next subsections, we will examine an appropriate method.

## 3.2 Wavelet Transform Basics

In this subsection we will revise general wavelet theory results necessary for the subsequent explanations. For a detailed information, refer to [22].

Let us call an array $s = (s_1, s_2, ..., s_n)$ of discrete values a signal. Let a high-pass wavelet filter be denoted as $h = (h_1, h_2, ..., h_t)$, and a low-pass wavelet filter be denoted as $l = (l_1, l_2, ..., l_t)$.

If to denote a convolution by $*$, and a dyadic downsampling by $_{\downarrow 2n}$, we can perform signal $s$ one-level wavelet decomposition as follows:

$$a_1 = s *_{\downarrow 2n} l; \ d_1 = s *_{\downarrow 2n} h. \tag{1}$$

In (1), $a_1$ is an array of approximation coefficients at level 1, whereas $d_1$ is an array of detail coefficients at the same level 1.

Also, we can apply (1) to $a_1$ receiving approximation and detail coefficients at level 2. In general, to obtain approximation and detail coefficients at any level $k$, we need to apply (1) to the approximation coefficients at level $k$-1:

$$a_k = a_{k-1} *_{\downarrow 2n} l = (\underbrace{(s *_{\downarrow 2n} l)...*_{\downarrow 2n} l}_{k \ times}); d_k = a_{k-1} *_{\downarrow 2n} h = (\underbrace{((s *_{\downarrow 2n} l)...*_{\downarrow 2n} l}_{k-1 \ times}) *_{\downarrow 2n} h) \ . \tag{2}$$

Any signal $s$ can always be presented as a following sum:

$$s = A_k + \sum_{u=1}^{k} D_u \ . \tag{3}$$

In (3), $A_k$ denotes an approximation at level $k$, and $D_u$ denotes a detail at level $u$. They can be obtained from the corresponding coefficients as follows:

$$A_k = (\underbrace{(a_k *_{\uparrow 2n} l)...*_{\uparrow 2n} l}_{k \ times}); \tag{4}$$

$$D_k = (\underbrace{((d_k *_{\uparrow 2n} h) *_{\uparrow 2n} l)...*_{\uparrow 2n} l}_{k-1 \ times}) \ . \tag{5}$$

In (4) and (5), $a_k$ and $d_k$ are being dyadically upsampled (which is denoted by $_{\uparrow 2n}$) first, and then they are convoluted with an appropriate wavelet filter.

When the length of $s$ is odd, performing dyadic downsampling becomes ambiguous. To get over this collision, we need to somehow make the signal length even. It can be done either by removing an element from the signal, or by adding a new one to it. Since removing elements always involves loss of data, adding a new sample is a lot more acceptable. In our opinion, extending the signal symmetrically (either leftwards or rightwards) seems to be the most adequate solution.

### 3.3    Modifying Approximations and Fixing Details

As we mentioned in Subsection 3.1, the approximation $A_k$ needs to be modified somehow. Coming from (**4**) and (**5**), the approximation depends only on the approximation coefficients, and the details depend only on the detail ones. Therefore, preserving the detail coefficients at level $k$ preserves the details at any level below $k$. Respectively, by modifying approximation coefficients at level $k$ we can modify the approximation at level $k$.

There exist two totally different approaches to transforming $A_k$. We called the first one an extremum transition approach. Applying it means performing such a modification that the resultant approximation's extremums totally differ from the initial ones. The other approach is called an "Ali Baba's wife" approach. Its name sends us back to the collection of the Middle Eastern and South Asian stories and folk tales "One Thousand and One Nights". One of the tales says that Ali Baba's wife marked all the houses in the neighborhood with the same symbol the thieves used to mark Ali Baba's house with. Having done that, she saved Ali Baba from an inevitable death. In terms of our paper, it means we do not eliminate existing extremums but add several alleged ones.

But, nobody can predict how changing the approximation coefficients will change the approximation without additional information. That's why we need to get such information about the signal.

### 3.4    Applying Wavelet Reconstruction Matrices to Modifying Approximations

It is known that WT can be performed using matrix multiplications [22]. In particular, we can rewrite (**4**) as follows:

$$A_k = M_{rec} \cdot a_k .    \textbf{(6)}$$

We will call $M_{rec}$ a wavelet reconstruction matrix (WRM). It can be obtained consequently multiplying appropriate upsampling and convolution matrices.

Now, let us apply WRM to solving our main task. As it was mentioned before, we need to find new approximation coefficients $\tilde{a}_k$. The structure of $M_{rec}$ always makes it possible to find appropriate solution (an illustrative example will be showed in the next section). After having chosen new coefficients, we can obtain a new approximation $\tilde{A}_k$ using (**6**). Then, we need to add $\tilde{A}_k$ and all the signal $c$ details. As a result, we get a new concentration signal $\tilde{c}$ .

According to Subsection 3.2, when the signal length is odd, we need to symmetrically extend it. In this case, it is necessary to ensure that the resultant signal $\tilde{c}$ is also a symmetric extension of any other odd-length signal. This means border signal elements have to be equal. We can always achieve that by fixing the difference between appropriate approximation values.

Besides, some $\tilde{c}$ elements can turn out to be negative. Since ratios cannot be negative, we have to make all $\tilde{c}$ elements positive (e.g., by adding to each of them a reasonably large value). But, in return we will get the signal with a completely different mean value. The only opportunity to overcome this problem is to multiply the signal by an appropriate value. Due to the algebraic properties of the convolution, both details' and approximation's elements will also be multiplied by the same value. This means the details will be changed proportionally, which totally suits our task definition requirements.

## 4   Experimental Results

To show the proposed algorithm in action and stress on its main features, we took the UK Census-2001 microfile provided by [3] as the data to analyze. The microfile contains information on more than 1,8 million respondents. For our sake, we decided to set a task of protecting the scientific professionals and technicians distribution over the regions of the UK. The importance of such a task is obvious. Maximums in an appropriate concentration signal can possibly lead to exposing the restricted scientific research centers which weren't supposed to be revealed. But, by adding alleged maximums to the signal we can guarantee that such centers will not be found out.

According to Subsection 3.1, we have to define both parameter and vital attributes and values. Since we intend to change regional distribution of scientists, we took "REGNUK" (which is an abbreviation of "Region of the UK") as a parameter attribute. Each value of this attribute stands for a particular region, making a total of 16 regions. Although, in the data extract provided by [3], there is no information on the "North", "East Anglia" and "Rest of South East" regions. Therefore, we were able to choose only 13 values left as parameter ones.

We also took "OCC" (which means "Occupation") as a vital attribute. This attribute values are three-digit numbers standing for various occupations and activities. But, since we're concerned in redistributing people of science only, we took just two vital values, i.e., "211" (for "Science Professionals") and "311" (for "Science and engineering technicians").

The next step is to build up a concentration signal. We counted up all the respondents with "Occupation" value "211" or "311" and a parameter value representing every region of the UK. These quantities are presented in Table 2 (the fourth row). Afterwards, we divided them by the overall number of employed people in each region.

We got the following concentration signal:

$c = (0.0143, 0.0129, 0.0122, 0.0140, 0.1149, 0.0141,$

$0.0142, 0.0128, 0.0077, 0.0100, 0.0159, 0.0168, 0.0110).$

In the paper, we present all the numeric data with 4 decimal numbers, but all the calculations were carried out with a higher proximity.

As we can see, the penultimate concentration is maximal. Further on, we will try to hide this maximum using "Ali Baba's wife" approach.

Since the resultant signal is of an odd length, we needed to add an additional element to it. We decided to symmetrically extend our signal leftwards.

Then, we used the second order Daubechies low-pass wavelet filter $l \equiv \left(\dfrac{1+\sqrt{3}}{4\sqrt{2}}, \dfrac{3+\sqrt{3}}{4\sqrt{2}}, \dfrac{3-\sqrt{3}}{4\sqrt{2}}, \dfrac{1-\sqrt{3}}{4\sqrt{2}}\right)$ to perform one-level wavelet decomposition (**2**) of $c$:

$$a_1 = (a_1(1), a_1(2), a_1(3), a_1(4), a_1(5), a_1(6), a_1(7)) =$$
$$= (0.0188, 0.0186, 0.0184, 0.0189, 0.0180, 0.0135, 0.0223).$$

The WRM for such a signal is as follows:

$$M_{rec} = \begin{pmatrix}
0.8365 & 0 & 0 & 0 & 0 & 0 & -0.1294 \\
0.2241 & 0.4830 & 0 & 0 & 0 & 0 & 0 \\
-0.1294 & 0.8365 & 0 & 0 & 0 & 0 & 0 \\
0 & 0.2241 & 0.4830 & 0 & 0 & 0 & 0 \\
0 & -0.1294 & 0.8365 & 0 & 0 & 0 & 0 \\
0 & 0 & 0.2241 & 0.4830 & 0 & 0 & 0 \\
0 & 0 & -0.1294 & 0.8365 & 0 & 0 & 0 \\
0 & 0 & 0 & 0.2241 & 0.4830 & 0 & 0 \\
0 & 0 & 0 & -0.1294 & 0.8365 & 0 & 0 \\
0 & 0 & 0 & 0 & 0.2241 & 0.4830 & 0 \\
0 & 0 & 0 & 0 & -0.1294 & 0.8365 & 0 \\
0 & 0 & 0 & 0 & 0 & 0.2241 & 0.4830 \\
0 & 0 & 0 & 0 & 0 & -0.1294 & 0.8365 \\
0.4830 & 0 & 0 & 0 & 0 & 0 & 0.2241
\end{pmatrix}.$$

According to (**6**), we obtained a signal approximation:

$A_1 = (0.0129, 0.0132, 0.0131, 0.0130, 0.0130, 0.0132, 0.0134, 0.0129,$
$0.0126, 0.0106, 0.0090, 0.0138, 0.0169, 0.0141).$

Also, we got a signal detail at level 1 according to (**5**):

$D_1 = (0.0014, 0.0011, -0.0003, -0.0008, 0.0010, -0.0017, 0.0007,$
$0.0013, 0.0002, -0.0029, 0.0011, 0.0021, -0.0007, -0.0031).$

To ensure that the difference between the first two approximation values will persist, we had to fix elements $a_1(1)$, $a_1(2)$, and $a_1(7)$, because the other 4 coefficients don't influence the first two approximation values when performing (**6**).

Taking into consideration the $M_{rec}$ elements in different rows, we can always pick such coefficients $\hat{a}_1$ that multiplication (**6**) yields different maximums in the resultant approximation. For example, if to take $\hat{a}_1 = (0.0188, 0.0186, -2, 0, 1, -5, 0.0223)$ we receive new maximal values in the 3^rd, the 9^th, and (as we intended to) the 13^th approximation elements. This particular choice isn't a unique one. In general, one can pick any other coefficients depending on the desired outcome.

So, in our case we got a new approximation:

$\hat{A}_1$ = (0.0129, 0.0132, 0.0131, -0.9618, -1.6754, -0.4483, 0.2588,

0.4830, 0.8365, -2.1907, -4.3120, -1.1099, 0.6657, 0.0141).

Having added old details to a new approximation, we got a new concentration signal:

$\hat{c} = \hat{A}_1 + D_1$ = (0.0143, 0.0143, 0.0129, -0.9626, -1.6744, -0.4500,

0.2595, 0.4843, 0.8367, -2.1935, -4.3109, -1.1078, 0.6656, 0.0110).

As we can see, some signal elements are negative. To make them all strictly positive, we added to each signal element 6.3109 (actually, we could add any other value large enough to make the signal positive):

$\bar{c}$ = (6.3252, 6.3252, 6.3238, 5.3484, 4.6365, 5.8609, 6.5704,

6.7952, 7.1476, 4.1174, 2.0000, 5.2031, 6.9765, 6.3220).

The only necessary condition we haven't met yet is the equality of corresponding mean values. For that sake, we multiplied $\bar{c}$ by a coefficient $\sum_{i=2}^{14} c_i / \sum_{i=2}^{14} \bar{c}_i = 0.0023$.

Here, we took into consideration only 13 last signal elements because the first one was added to make signal length even, and doesn't contain any necessary information.

The resultant signal is presented in Table 2 (the last row).

The last step is to obtain new quantities. We have done that by multiplying new ratios by a total number of employed people in each region. As quantities can be only integers, we had to round them afterwards (see Table 2, the sixth row).

**Table 2.** Quantities and ratios distributed by regions.

| Column number | 1 | 2 | 3 | 4 | 5 | 6 | 7 |
|---|---|---|---|---|---|---|---|
| Region code | 11 | 13 | 14 | 21 | 22 | 31 | 33 |
| Employed | 48591 | 129808 | 96152 | 83085 | 101891 | 108120 | 161395 |
| Scientists (initial) | 695 | 1672 | 1176 | 1163 | 1171 | 1524 | 2294 |
| Signal $c$ (initial) | 0.0143 | 0.0129 | 0.0122 | 0.0140 | 0.1149 | 0.0141 | 0.0142 |
| Scientists (final) | 699 | 1867 | 1170 | 876 | 1358 | 1616 | 2495 |
| Signal $c$ (final) | 0.0144 | 0.0144 | 0.0122 | 0.0105 | 0.0133 | 0.0149 | 0.0155 |

| Column number | 8 | 9 | 10 | 11 | 12 | 13 | Mean |
|---|---|---|---|---|---|---|---|
| Region code | 40 | 51 | 52 | 60 | 70 | 80 | |
| Employed | 97312 | 54861 | 86726 | 99890 | 55286 | 33409 | |
| Scientists (initial) | 1246 | 422 | 871 | 1589 | 927 | 369 | 1163 |
| Signal $c$ (initial) | 0.0128 | 0.0077 | 0.0100 | 0.0159 | 0.0168 | 0.0110 | 0.0129 |
| Scientists (final) | 1582 | 514 | 395 | 1182 | 877 | 480 | 1162.4 |
| Signal $c$ (final) | 0.0163 | 0.0094 | 0.0045 | 0.0118 | 0.0159 | 0.0144 | 0.0129 |

Though the resultant data completely differ from the initial ones, we preserved both mean value and wavelet decomposition details.

It is important to note that rounding the quantities may lead to some changes in wavelet decomposition details. Though, in most cases they are not very significant.

All that is left to fulfil is to construct a new microfile. We can always do that by changing vital values of different records according to the received quantities.

It is obvious that picking different vital and parameter attributes, or even different WRMs, doesn't restrict the possibility of applying the method under review to providing group anonymity.

## 5 Conclusion and Future Research

In the paper, we attracted attention to the problem of providing group anonymity while preparing microdata. In response to this new challenge, we introduced a totally novel wavelet-based method for providing group anonymity in collective data.

It is significant to state that the proposed method might be combined with those for providing individual anonymity without any restrictions. Thus, it can be implemented in a real-life privacy-preserving data mining system.

We beleive that the current paper cannot suggest answers to all the questions and problems arising. In our opinion, there exist many other kinds of group anonymity tasks to study in the future.

Apart from it, we can distinguish the following problems:

- It is important to introduce group anonymity measure.
- Using different wavelet bases leads to different WRMs, so it is interesting to study the opportunities they provide when modifying approximation coefficients.

## References

1. Gantz, J.F., Reinsel, D.: As the Economy Contracts, the Digital Universe Expands. An IDC Multimedia White Paper (2009),
   http://www.emc.com/collateral/demos/microsites/idc-digital-universe/iview.htm
2. Mullins, I., Siadaty, M., Lyman, J., Scully, K., Garrett, C., Miller, W., Muller, R., Robson, B., Apte, C., Weiss, S., Rigoutsos, I., Platt, D., Cohen, S., Knaus, W.: Data Mining and Clinical Data Repositories: Insights from a 667,000 Patient Data Set. Computers in Biology and Medicine, 36(12), 1351--1377 (2006)
3. Minnesota Population Center. Integrated Public Use Microdata Series International, https://international.ipums.org/international/
4. Health Insurance Portability and Accountability Act of 1996 (HIPAA). Public Law 104-191, 104th Congress, Aug. 21, 1996, http://www.hipaa.org/
5. Patient Safety and Quality Improvement Act of 2005 (PSQIA). Federal Register, 73(266), 2001
6. Directive 2002/58/EC of the European Parliament and of the Council of 12 July 2002. Official Journal of the European Communities, L 201, 37--47, 31/07/2002
7. Chertov, O., Pilipyuk, A.: Statistical Disclosure Control Methods for Microdata. In: International Symposium on Computing, Communication and Control, pp. 338--342. IACSIT, Singapore (2009)
8. Agrawal, R., Srikant, R.: Privacy-Preserving Data Mining. In: ACM SIGMOD International Conference on Management of Data, pp. 439--450. ACM Press, Dallas, Texas (2000)

9. Domingo-Ferrer, J., Mateo-Sanz, J.M. Practical Data-oriented Microaggregation for Statistical Disclosure Control. IEEE Transactions on Knowledge and Data Engineering, 14(1), 189--201 (2002)

10. Fienberg, S., McIntyre, J.: Data Swapping: Variations on a Theme by Dalenius and Reiss. Technical Report, National Institute of Statistical Sciences (2003)

11. Domingo-Ferrer, J.: A Survey of Inference Control Methods for Privacy-Preserving Data Mining. In: Aggarwal, C.C., Yu, P.S. (eds.): Privacy-Preserving Data Mining: Models and Algorithms, pp. 53--80. Springer, New York (2008)

12. Xu, S., Zhang, J., Han, D., Wang, J.: Singular Value Decomposition Based Data Distortion Strategy for Privacy Protection. Knowledge and Information Systems, 10(3), 383--397 (2006)

13. Wang, J., Zhong, W.J., Zhang, J.: NNMF-based Factorization Techniques for High-accuracy Privacy Protection on Non-negative-valued Datasets. In: IEEE Conference on Data Mining, International Workshop on Privacy Aspects of Date Mining, pp. 513--517. IEEE Computer Society, Washington (2006)

14. Liu, L., Wang, J., Lin, Z., Zhang, J.: Wavelet-Based Data Distortion for Privacy-Preserving Collaborative Analysis. Technical Report No. 482-07, Department of Computer Science, University of Kentucky, Lexington (2007)

15. Davydov, A.: Wavelet-analysis of the Social Processes. Sotsiologicheskie issledovaniya, 11, 89--101 (2003) (in Russian),
http://www.ecsocman.edu.ru/images/pubs/2007/10/30/0000315095/012.DAVYDOV.pdf

16. The Free Haven Project, http://freehaven.net/anonbib/full/date.html

17. Evfimievski, A.: Randomization in Privacy Preserving Data Mining. ACM SIGKDD Explorations Newsletter, 4(2), 43--48 (2002)

18. Kargupta, H., Datta, S., Wang, Q., Sivakumar, K.: Random-data Perturbation Techniques and Privacy-preserving Data Mining. Knowledge and Information Systems, 7(4), 387--414 (2005)

19. Aggarwal, C.C.: On Randomization, Public Information and the Curse of Dimensionality. In: 23rd International Conference on Data Enginering, pp. 136--145. IEEE Computer Society, Washington (2007)

20. Sweeney, L.: *k*-anonymity: a Model for Protecting Privacy. International Journal on Uncertainty, Fuzziness and Knowledge-based Systems, 10(5), 557--570 (2002)

21. Mallat, S.: A Wavelet Tour of Signal Processing. Academic Press, New York (1999)

22. Strang, G., Nguyen, T.: Wavelet and Filter Banks. Wellesley-Cambridge Press, Wellesley (1997)

23. Li, T., Li, Q., Zhu, S., Ogihara, M.: A Survey on Wavelet Applications in Data Mining. ACM SIGKDD Explorations Newsletter, 4(2), 49--68 (2002)

24. Bapna, S., Gangopadhyay, A.: A Wavelet-based Approach to Preserve Privacy for Classification Mining. Decision Sciences Journal, 37(4), 623--642 (2006)

25. Liu, L., Wang, J., Zhang, J.: Wavelet-based Data Perturbation for Simultaneous Privacy-Preserving and Statistics-Preserving. In: 2008 IEEE International Conference on Data Mining Workshops, pp. 27--35. IEEE Computer Society, Washington (2008)